\documentclass{amsart}

\usepackage[T1]{fontenc}

\usepackage[foot]{amsaddr}
\usepackage{graphicx}
\usepackage{subfig}
\usepackage{tikz}
\usepackage{tkz-fct}
\usetikzlibrary{shapes,shadows,arrows,positioning,patterns,shadows.blur,math}
\newcommand*\circled[1]{\tikz[baseline=(char.base)]{\node[shape=circle,draw,inner sep=2pt] (char) {#1};}}
\begin{document}

\title{Airborne Virus Transmission: Increased Spreading due to Formation of Hollow Particles}

\author{Gizem Ozler$^{1,2,*}$}
\email{gizem.oezler@ptb.de}

\author{Holger Grosshans$^{1,2}$}
\email{holger.grosshans@ptb.de}

\address{$^1$ Physikalisch- Technische Bundesanstalt (PTB), Braunschweig 38116, Germany}
\address{$^2$ Otto von Guericke University of Magdeburg, Institute of Aparatus and Environmental Technology, Magdeburg 39106, Germany}


\begin{abstract}
The globally supported social distancing rules to prevent airborne transmission of COVID-19 assume small saliva droplets evaporate fast and large ones, which contain most viral copies, fall fast to the ground.
However, during evaporation, solutes distribute non-uniformly within the droplets.
We developed a numerical model to predict saliva droplet drying in different environments.
We represent saliva droplets as a solution of NaCl mixed with water.
In a hot and dry ambiance, the solutes form a shell on the droplets' surface, producing light, hollow particles.
These hollow particles have a larger cross-sectional area compared to their solid counterparts and can float longer and travel farther in the air.
We introduced the "hollowness factor," which serves as a measure of the ratio of the volume of a hollow particle and the volume of a solid residue formed during droplet drying. Through our investigations, we determined that under specific conditions, namely an ambient humidity level of 10\% and a temperature of 40~°C, the highest hollowness factor observed was 1.610. This finding indicates that in the case of hollow particle formation, the droplet nucleus expands by a factor of 1.610 compared to its original size.

\end{abstract}

\maketitle

\section{Introduction}

Starting in March 2020, the world faces the challenge of controlling the Coronavirus (COVID-19) outbreak. 
Infection occurs via virus-containing saliva droplets exhaled from human respiratory systems during, for example, breathing, speaking, or coughing \cite{WHO}.
Besides vaccinations, preventing virus transmission is the key to stopping the spread of the disease.
To this end, tremendous research aimed to understand the transmission mechanisms of SARS-CoV-2, the cause of the Coronavirus disease.
Moreover, governments imposed omnipresent preventive measures, such as facial masks, social distancing, and lockdowns.
Despite all that, high infection rates repeatedly demonstrate how little we know about virus transmission.

Viruses transmit either directly or indirectly.
Directly through physical contact between an infected and a susceptible person.
Indirectly, either through contact with contaminated surfaces (also called fomites) or airborne by respiratory droplets and aerosols \cite{Brankston}.
Physical contact is prevented by social distancing, and fomites pose a low risk of transmitting the disease \cite{Goldman, Lewis}.
Thus, the least controllable and most dangerous transmission mode is airborne \cite{Morawska, Zhang}.

The trajectories of airborne saliva droplets depend on the aerodynamic and gravitational forces acting on them and, thus, on the droplets' sizes.
In addition, once a droplet is exhaled, it evaporates and decreases in size. 
In other words, a droplet’s trajectory is not determined by its initial size but by its evolution. 
Therefore, droplet evaporation plays a crucial role in airborne virus transmission.

In his classical 1934 work \cite{Wells}, Wells predicted the lifetime of evaporating pure water droplets using the well-known D-2 law.
According to the D-2 law, the droplets' surface decreases at a constant rate depending on their composition and the ambient conditions.
Following Wells, evaporation limits the lifetime of small droplets emitted from the human respiratory system, whereas large droplets fall to the ground.
Both limiting mechanisms coincide at a droplet size of 140~µm.
Consequently, droplets of a size of 140~µm live the longest, namely around 3~s. 
Within their lifetime, these droplets are expected not to propagate further than 1-2 meters \cite{WHO2}.

These principle thoughts justify social distancing up to today.
According to clinicians and medical scientists, particles smaller than 5 µm can penetrate deep into the lungs \cite{Thakur}.
However, due to their low inertia and short evaporation time, these small particles remain in the surrounding of an infected person and pose little risk for airborne virus transmission \cite{Ahmadzadeh}.
Due to their high inertia, clinicians assume droplets larger than 5 µm rapidly fall to the ground within 1-2 meters.
Public health agencies endorse this argument in the globally promoted social distancing rule.

Given the all-embracing consequences of the distancing rule on most of the world population, it is surprising what a simple evaporation model it bases on.
The above-mentioned prediction of Wells assumes the evaporation of pure water droplets.
However, saliva contains multiple non-volatile components like salts, proteins, and other organic and inorganic substances \cite{Almeida}.
These non-volatile components reduce the vapor pressure and, thus, the evaporation rate.
Also, after the volatile components evaporate, the non-volatile components form droplet nuclei.
The formation of such nuclei after evaporation was proven in laboratory experiments using mechanical suspensions \cite{Czerwiec}, optical trapping devices \cite{Liu}, and acoustic levitation of saline \cite{Chaudhuri, Basu} and saliva droplets \cite{Stiti, Lieber}.
These nuclei are lighter than the initial droplets and travel longer distances by air, greatly increasing the risk of infection \cite{Jayaweera}.
There is increasing evidence that distancing by 1-2 meters protects insufficiently; 
respiratory particles can travel airborne over longer distances \cite{Bourouiba, Xie}, and particles up to 100 µm can remain suspended in the air \cite{Speight}.

While most experiments observe the transformation of droplets into nuclei from the outside, numerical simulations can extract detailed internal data.
However, simulations of saliva droplet drying usually adopt a zero-dimensional approach \cite{Chaudhuri, Stiti, Ugarte-Anero, Chillon, Biswas}.
This approach implies that all components are always homogeneously distributed within the droplet's volume.
Also, it implies that the droplet's equilibrium diameter is given a priori by the constant solid content and the mixture's efflorescence limit.

\begin{figure}[tb]
\centering
\includegraphics[width=10cm]{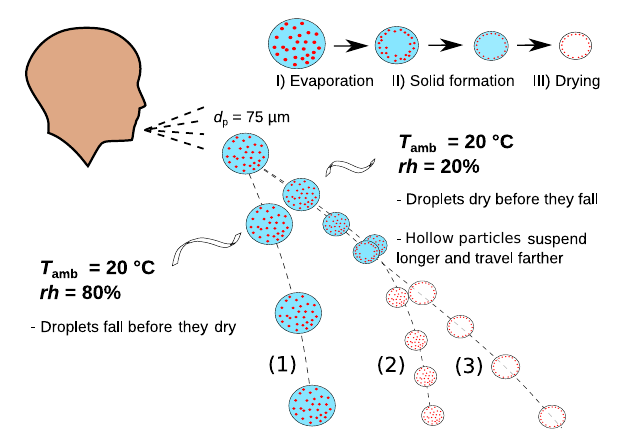}
\caption{
Often, large saliva droplets fall fast to the ground (1).
However, in certain weather conditions, nuclei form that travel much further (2).
These nuclei can form hollow particles due to surface enrichment, and these hollow particles might travel even further (3).
}
\label{covid}

\end{figure}

Contradicting the one-dimensional simulation approach, experiments revealed that during evaporation of salt-water droplets, which closely resembles saliva, the salt is not homogeneously distributed \cite{Gregson}.
Instead, evaporation enriches the surface with solutes, thereby competing against internal diffusion.
Due to surface enrichment, a droplet effloresces at its surface faster than a perfectly mixed droplet.
Once the droplet effloresces at the surface, a solid layer forms that prevent further shrinking.
In that case, the resulting particle is larger than predicted by assuming homogeneity \cite{Rezaei}.

The final particle morphology is significantly influenced by the pressure build-up inside  the drying droplet. Following the formation of a solid layer, the size of the droplet remains unchanged, while the liquid inside evaporates by passing through the pores of the crust. However, the evaporation process encounters growing resistance, resulting in an elevation of the droplet's temperature. Eventually, the droplet temperature reaches the same level as the surrounding air temperature. If the air temperature is higher than the boiling point of the liquid mixture, vapor accumulates within the particle. Since the accumulated vapor cannot easily escape through the pores, this may lead to an increase in pressure inside the particle. Depending on the mechanical properties of the solid layer, the particle may undergo cracking or explosion \cite{Nesic}. 
Capillary pressure is also a significant factor in determining the ultimate particle morphology. Droplets shrinking results in an elevated capillary pressure due to the reduced radius. When crust formation prevents further droplet shrinkage, solvent evaporation persists through the porous crust, leading to a continued increase in the pressure difference across the crust. The mechanical characteristics of the crust determine its ability to withstand this pressure build-up, ultimately influencing the resulting particle morphology.
When the pressure difference exceeds the threshold at which the strength of the crust cannot withstand, buckled or collapsed particles can occur \cite{Boel}.
Our research does not focus on investigating the impact of capillary forces, and boiling of the liquid mixture is not relevant since the ambient air temperature is far less than the boiling temperature of the liquid mixture. 
Consequently, we expect a hollow particle structure to form once all the water content has fully evaporated. Multiple experimental studies have provided compelling evidence, supported by SEM images, demonstrating the formation of hollow particles resulting from the drying of bi-component droplets. These studies encompass various mixtures, including NaCl-water\cite{Walton, Hardy, Zhang, Vehring}, among others\cite{Vicente, Lucas}.

Hollow particles may dramatically change their dynamics compared to solid ones.
Aerodynamic forces increase due to their larger cross-sectional area, but gravitational forces remain constant due to the equal solid mass.
In other words, the local particle Stokes number reduces.
Consequently, an atmospheric flow can transport a hollow particle over larger distances than its solid counterpart.

The degree of hollowness depends on the competing rates of solute diffusion within the droplet and evaporation at the droplet's surface.
Since the sensitivities of evaporation and diffusion on the drying conditions are different, the weather conditions alter the balance of both rates.
Thus, we suspect the accepted social distancing rule holds for some weather conditions.
However, for some surrounding temperature and humidity combinations, saliva droplets might form hollow particles that travel much longer distances than expected and predicted by current models (see Fig.\ref{covid}).

In essence, predicting the drying kinetics requires an unsteady and at least one-dimensional analysis.
The equilibrium diameter is unknown a priori.
Therefore, considering inhomogeneous component distribution, we developed a new model for the morphological evolution of saliva droplets during evaporation.
The saliva in the model is represented by a solution of water and NaCl.
We modeled the formation of a solid shell once a droplet reaches its efflorescence limit.
After this point, the particle size remains fixed, and the water inside the shell dries out.

Our study aims to determine whether the formation of a hollow particle can change our perception of airborne virus transmission.
We identify the relevant weather conditions and droplet size classes particularly prone to hollow particles.
Then, we estimate the increase in the traveling distance due to hollow particle formation. 

The weather conditions affect not only droplet evaporation and their trajectories but also the infectivity of the airborne Coronavirus \cite{Oswin2021}.
Moreover, the initial droplet diameter also determines the virus load \cite{Wang_virus}.
Therefore, we relate our results regarding hollow particles to the expected virus infectivity and load to receive an overall picture of the danger of certain weather conditions and droplet sizes.

\section{Methods}
\subsection{Mathematical Model}

We modeled the evaporation, drying, solid layer formation, and transport of single, spherical saliva droplets. 
We modeled saliva droplets as a bicomponent salt(NaCl)-water mixture, and we used the initial concentration of 6 g NaCl/kg liquid mixture provided by Stiti et al. \cite{Stiti}. The droplets were simulated for different initial sizes, ambient temperatures, and relative humidities to assess the evaporation dynamics in different ambients. 

The droplet evaporation model expands the model of Gopireddy and Gutheil \cite{Gopireddy} for spray dryers to the drying of saliva after a cough, including surface enrichment and particle formation. 
We expanded the model of Gopireddy and Gutheil \cite{Gopireddy} by including the spatial changes in diffusivity coefficient and thermal diffusivity based on the local concentration and temperature. We considered the effect of the cough and ambient velocity field on the droplet evaporation rate. Additionally, the properties of the NaCl-water mixture have been adapted to the existing model.
Calculations of the physical and transport properties of the liquid mixture are given in supplementary information.

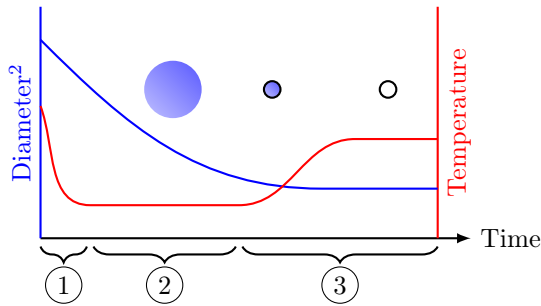
\begin{figure}[tb]
\centering
\begin{tikzpicture}[thick,scale=0.22]
\draw[->,>=latex] (0,0) -- (26,0) node[right] {Time};
\draw[blue] (0,0) -- (0,14.0) node[midway,rotate=90,above,blue] {Diameter$^2$};
\draw[red] (24,0) -- (24,14.0) node[midway,rotate=90,below,red] {Temperature};
\draw[blue] (0,12) to [out=-45,in=180] (17,3) to [out=0,in=180] (24,3);
\draw[red] (0,8) to [out=-70,in=180] (3,2) to [out=0,in=180] (12,2) to [out=0,in=180] (19,6) to (24,6);
\draw[white,shading = axis,rectangle, left color=blue!30!white, right color=blue!60!white,shading angle=135] (8,9) circle [radius=1.8];
\draw[shading = axis,rectangle, left color=blue!30!white, right color=blue!60!white,shading angle=135] (14,9) circle [radius=.5];
\draw[] (21,9) circle [radius=.5];
\draw[thick,decorate,decoration={mirror,brace,amplitude=6pt}] (0,-.5) -- (2.8,-.5) node[midway,below,yshift=-4pt,]{\circled{1}};
\draw[thick,decorate,decoration={mirror,brace,amplitude=6pt}] (3.2,-.5) -- (11.8,-.5) node[midway,below,yshift=-4pt,]{\circled{2}};
\draw[thick,decorate,decoration={mirror,brace,amplitude=6pt}] (12.2,-.5) -- (24,-.5) node[midway,below,yshift=-4pt,]{\circled{3}};
\end{tikzpicture}
\caption{Schematics of our new saliva drying model.
During stage 1, the droplet cools from body temperature close to the surrounding air’s wet-bulb temperature.
Stage~2 is the quasi-equilibrium evaporation, the non-evaporating solutes reduce the evaporation rate.
In stage~3, solutes form a solid layer on the surface, liquid further evaporates, and finally, the particle reaches the surrounding air temperature.
}
\label{stages}
\end{figure}

Our model distinguishes three stages of saliva drying, as sketched in Fig.~\ref{stages}.
During the first stage, the droplets cool down from body temperature close to the wet-bulb temperature of the surrounding air with some solvent evaporation.
In the second stage, the droplets evaporate in a quasi-equilibrium state.
With the evaporation of the solvent, the droplets become smaller, and solute concentration increases.
The evaporation rate decreases with increasing solute concentration at the droplets' surface.
The third stage begins when the solute concentration at the surface reaches the efflorescence limit, and on the droplet's surface, a solid layer begins to form.
The droplet size stays constant from that point on, but the solid layer's thickness increases with drying.
The temperature of the droplet increases as the evaporation rate decreases during this period. 
Finally, drying occurs through the solid layer, forming a droplet nucleus.
Depending on when the droplet effloresces, a hollow particle can form.

For the species mass and heat conservation, 1-D diffusion equations in radial coordinates for a spherically symmetric droplet are solved.
The equation for the concentration of the species is given by
\begin{equation}
\frac{\partial Y_{\text {i}}}{\partial t}=\frac{D_{\text {12}}}{r^{2}}\left[  \frac{\partial }{\partial r}\left( r^{2}\frac{\partial Y_{\text {i}}}{\partial r} \right)\right],
\label{eq1}
\end{equation}
where $Y_{\text {i}}$ is the mass fraction of the component $i$ (i = 2; 1 = water, 2 = NaCl), $r$ is the radial coordinate, $D_{\text {12}}$ is the binary diffusion coefficient of salt-water, and $t$ is the time. 
We expanded the model of Gopireddy and Gutheil \cite{Gopireddy} by solving, in each time step, the binary diffusion coefficient of NaCl-water mixture as a function of concentration and temperature for each grid node within the droplet.
Furthermore, evaporative cooling is considered when calculating this property.
In order to account for concentration and temperature-dependent salt-water diffusivity, we implemented the relation reported by Gregson et al. \cite{Gregson}.

We assumed that, initially, the droplet is a homogeneous mixture, i.e., $Y_{\text {i}}=Y_{\text {i,0}}$ for $t=0$.
The regularity condition is satisfied at the center of the droplet, $r=0$, $\partial Y_{\text {i}}/{\partial r}=0$.
The boundary condition to Eq.~(\ref{eq1}) at the droplet surface, $r = R(t)$,  considers the change in the droplet size,
\begin{equation}
-D_{\text {12}}\frac{\partial Y_{\text {i}}}{\partial r}-Y_{\text {i}}\frac{\partial R}{\partial t}=\frac{\dot{m}}{A \rho_\mathrm{l}}.
\label{eq2}
\end{equation}
Here, $R$ and $A$ are the radius and the surface area of the droplet at a given time, $\dot{m}$ is the evaporation rate, $\rho_{\text {l}}$ is the density of droplet.

The calculation of the evaporation rate (Eq.~(\ref{eq2})) requires several assumptions, described in the following. The model assumes that the droplet maintains a spherical shape throughout the evaporation process. The spherical symmetry is assumed for the droplet, temperature, and vapor mass concentration distributions in the surrounding gas. In the context of heat and mass transfer in an evaporating droplet, the quasi-steady approach is employed for the gas phase. This assumption is justified by the higher density and consequently higher inertia of the liquid phase. As a result, the properties of the liquid phase change at a much slower rate compared to the gas phase properties \cite{Law}. That means the rate of reduction of the size of the droplet is much smaller than the magnitude of vapor phase velocity. Thus, the radius, concentration, and temperature of the droplet can be assumed constant solving the evaporation rate. In other words, the gas phase properties adapt directly to the new boundary conditions. Consequently, the evaporation rate is \cite{Abramzon}
\begin{equation}
\dot{m}=2\pi R_\mathrm{eq}\,\rho_\mathrm{f}\,D_\mathrm{f}\,\widetilde{\textup{Sh}}\ln(1+B_{\text {M}}).
\label{eq3}
\end{equation}
Here, $R_{\text {eq}}$ is the volume equivalent diameter, $R_{\text {eq}}=(V_{\text {i}}/V)^{1/3}$, $\rho_{\text {f}}$ and $D_{\text {f}}$ are the density and the diffusivity of the film, $\widetilde{\textup{Sh}}$ is the modified Sherwood number to take into account the effect of convective droplet evaporation, and $B_{\text {M}}$ is the Spalding mass transfer number given in Eq.~(\ref{eq4}). Definitions of the modified mass and heat transfer coefficients are given in supplementary information.
\begin{equation}
B_{\text {M}} = \frac{Y_\mathrm{v,s}-Y_\mathrm{v,\infty }}{1-Y_\mathrm{v,s}}.
\label{eq4}
\end{equation}
The vapor mass fractions at the droplet's surface and in the ambient are $Y_\mathrm{v,s}$ and $Y_\mathrm{v,\infty}$ respectively. 
$Y_\mathrm{v,s}$ is calculated assuming a phase equilibrium at the droplet's surface. Non-ideal behavior by the presence of solute is taken into account by the activity coefficient. The supplementary information to this paper provides calculations for the activity coefficient and vapor mass fraction of the droplet's surface.

In the calculation of film properties, we used the reference temperature given as~\cite{Hubbard} 
\begin{equation}
T_{\textup{r}}=2/3\,T_\textup{s}+1/3\,T_{\textup{amb}}.
\label{eq5}
\end{equation}
Here, $T_{\textup{s}}$ is the droplet surface temperature, and $T_{\textup{amb}}$ is the ambient temperature.

The temperature profiles within a droplet are computed from the droplet's conductive heat transfer,
\begin{equation}
\frac{\partial T}{\partial t}=\frac{\alpha}{r^{2}}\left[  \frac{\partial }{\partial r}\left( r^{2}\frac{\partial T}{\partial r} \right)\right],
\label{eq6}
\end{equation}
where $T$ is the liquid mixture's temperature and $\alpha$ is the thermal diffusivity.
Compared to the simulations of Gopireddy and Gutheil \cite{Gopireddy}, we solved the thermal diffusivity in each grid cell updating thermal conductivity, specific heat capacity, and liquid mixture density as a function of local temperature and concentration.
Initially, the droplet has a uniform temperature of $T=T_0$ for $t=0$.
At the center of the droplet, $r=0$, zero gradient condition is applied, $\frac{\partial T}{\partial r}=0$. 
At the droplet surface, the energy balance reads
\begin{equation}
k_{\textup{l}}\frac{\partial T}{\partial r}=h_\textup{f}(T_{\textup{amb}}-T_{\textup{s}})+L_{\textup{v}}(T_{\textup{s}})\,\rho_{\textup{l}}\,\frac{\partial R}{\partial t}.
\label{eq7}
\end{equation}
Here, $k_{\textup{l}}$ is the thermal conductivity of liquid mixture, $h_\textup{f}$ is the heat transfer coefficient of the film, $L_{\textup{v}}(T_{\textup{s}})$ is the latent heat of vaporization at $T_{\textup{s}}$, $\rho_{\textup{l}}$ is the density of the liquid mixture.

The droplet evaporation rate is affected by the cooling of the droplet due to evaporation, reducing the temperature of the droplet and, thus, the water vapor concentration.
Therefore, the evaporative cooling of droplets is taken into account in our model.
In addition, non-volatile solutes are considered in the evaporative cooling process.

When the mass fraction of NaCl at the surface of the droplet reaches the efflorescence limit, a spherical solid layer forms and the droplet radius remains constant. Gregson et al. \cite{Gregson}, performed experiments on the drying of single NaCl-water droplets which are levitated by electrodynamic balance. Throughout their experiments, the drying process of the droplets is monitored until efflorescence is achieved. This occurrence of efflorescence induces noticeable alterations in both the concentration and shape of the droplets. Consequently, it becomes possible to measure the efflorescence time and, subsequently, determine the corresponding efflorescence concentration. The efflorescence limit is offered as $Y_{\textup{2,s}} = 0.393$. During the drying process, the solid layer at the surface thickens.
The drying rate due to solid layer resistance is computed by
\begin{equation}
\dot{m}=\frac{2\pi R_\mathrm{eq}\,\rho_\mathrm{f}\,D_\mathrm{f}\,\widetilde{\mathrm{Sh}}\ln(1+B_{\text {M}})}{1+\widetilde{\mathrm{Sh}}\,D_\mathrm{f}\,\delta\,/\,[\,2D_\mathrm{s}\,(R-\delta)\,] },
\label{eq8}
\end{equation}
where $\delta$ is the solid layer thickness, $D_\mathrm{s}$ is the diffusivity in the solid layer.  Our research does not examine the effect of capillary force on water vapor diffusion through solid layers.

 The temperature gradients are very small during the first and second stages of evaporation, with Biot number ($\textup{Bi}={\widetilde{\textup{Nu}}}\,k_\textup{f}\,/2.0\,k_\textup{l}$) less than 0.5. Therefore, we assumed uniform temperature of droplets after solid layer formation and the consequent temperature evolution is solved by

\begin{equation}
mC_{\textup{p}_\textup{l}}\frac{d\,T_\textup{s}}{dt}=\frac{2\pi R_\mathrm{eq}\,k_\mathrm{f}\,\widetilde{\textup{Nu}}(T_{\textup{amb}}-T_{\textup{s}})\ln(1+B_{\text {T}})}{{1+\widetilde{\textup{Nu}}\,k_\mathrm{\,f}\,\delta\,/\,[\,2k_\mathrm{s}\,(R-\delta)\,] }}-\dot{m}L_{\textup{v}} (T_\textup{s}),
\label{eq9}
\end{equation}
where $k_\mathrm{\,f}$ is the thermal conductivity of film, $\widetilde{\textup{Nu}}$ is the modified Nusselt number to consider the effect of convective droplet evaporation, and $k_\mathrm{\,s}$ is the thermal conductivity of the solid layer, and $B_{\text {T}}$ is the Spalding heat transfer number given as

\begin{equation}
\begin{aligned}
B_{\text {T}}& =(1+B_{\text {M}})^{\phi}-1, \\
\phi&= (C_{\textup{p,l}}/C_{\textup{p,f}})(\widetilde{\textup{Sh}}/\widetilde{\textup{Nu}})(1/\textup{Le}).
\label{eq10}
\end{aligned}
\end{equation}
Here, $C_{\textup{p,l}}$ is the specific heat capacity of the liquid mixture, $C_{\textup{p,f}}$ is the specific heat capacity of the film, and Le is the Lewis number.

Following the model of Gopireddy and Gutheil \cite{Gopireddy2}, it is assumed that the crust diffusion coefficient is $D_\mathrm{s}=2\,\widetilde{\mathrm{Sh}}\,D_\mathrm{f}$ and the thermal conductivity of the solid layer is $k_\mathrm{s}=10\,k_\mathrm{\,f}$. The premise rests on the fact that the crust diffusion coefficient is generally greater than the convection-diffusion coefficient of the droplet's boundary layer, $\widetilde{\textup{Sh}}\,D_{\text {f}}$. Furthermore, the ratio of heat diffusion coefficients ($k_\mathrm{\,f}/k_\mathrm{s})$ is significantly smaller than the ratio of mass diffusion coefficients ($D_\mathrm{\,f}/D_\mathrm{s})$ \cite{Nesic, Gopireddy2}.

We simulated the transport of droplets emitted from a cough to estimate the falling time and the maximum traveling distance.
Calculating the saliva droplet velocity requires superimposing both the cough-generated and ambient flows.
We used the velocity field measured by Wang et al. \cite{Wang_cough} through Particle Image Velocimetry (PIV) for the coughed airflow.
Accordingly, the distance of the cough-flow from the month, $s$, varies as $t^{0.3}$ and the relation between the cough's convection velocity, $u_\mathrm{c,max}$, and $s$ is
\begin{equation}
u_{\textup{c,max}}=\left\{\begin{matrix}
6.48~~\mathrm{m/s} & s\leq 0.08~\mathrm{m}\\ 
0.3(s+0.188)^{-7/3}~~\mathrm{m/s} & s\geq 0.08~\mathrm{m}
\end{matrix}\right..
\label{eq11}
\end{equation} 
The velocity of cough with respect to the vertical coordinate, $y$, is approximated as a hyperbolic cosine function related to the maximum velocity at the center
\begin{equation}
u_\textup{c}/u_{\textup{c,max}}=\textup{cosh}^{-2}(y/b_{\textup{1/2}})~,
\label{eq12}
\end{equation} 
where $b_{\textup{1/2}}$ is the jet half-width.

If the droplet is outside of the boundary of the cough airflow, the velocity profile is defined by
\begin{equation}
u_\textup{f} = [(2\,y/\,1.8~\mathrm{m}) - (y\,/\,1.8~\mathrm{m})^2]\,0.3~\mathrm{m/s}\,.
\label{eq13}
\end{equation}
Here, we used a parabolic function to define a velocity field that resembles an indoor flow.
The maximum velocity of the flow is 0.3 m/s \cite{Gong} at the average human height (1.8 m).
At the floor, $y=0$, the above equation satisfies the no-slip condition.

If the droplet is inside the cough, the horizontal component of the air velocity is the sum of $u_\textup{c}$ and $u_\textup{f}$. 
We assumed that the vertical component of the air velocity is zero.

Newton's second law of motion is used to obtain droplet's trajectories:
\begin{equation}
m_{\textup{d}}\frac{d{\textbf{\textit{u}}}_\text d}{dt}=-\frac{\pi }{8}\rho_{\textup{g}}C_{\textup{d~}}d^{2}\left \| \textbf{\textit{u}}_{\text d}-\textbf{\textit{u}} \right \|\left (\textbf{\textit{u}}_{\text d}-\textbf{\textit{u}} \right )+m_{\textup{d~}}\textbf{\textit{g}}.
\label{eq14}
\end{equation}
Here, $m_{\textup{d}}$ is the mass of droplet, $\textbf{\textit{u}}_\text d$ is the droplet velocity, $\textbf{\textit{u}}$ is the total air velocity, $d$ is the droplet diameter, and $C_{\textup{d~}}$ is the drag force coefficient computed as \cite{Putnam}
\begin{equation}
{C}_{\textup{d}}=\left\{\begin{matrix}
24(1+0.15 \, \textup{Re}_\textup{d}^{0.687})/\textup{Re}_\textup{d}&~\textup{Re}_\textup{d}\leq 1000 \\ 
 0.44& ~\textup{Re}_\textup{d}>  1000
\end{matrix}\right.
\label{eq15}
\end{equation}
where $\textup{Re}_{\textup{d}}$ is the droplet Reynolds number, $\textup{Re}_{\textup{d}} = 2\,\rho_\textup{g}\,u_{\textup{rel}}R/\mu_\textup{f}$, as a function of relative velocity of droplet $u_{\textup{rel}}$.

\subsection{Validation of Model}

\begin{figure}[h]
    \centering
        \includegraphics[width=8cm]{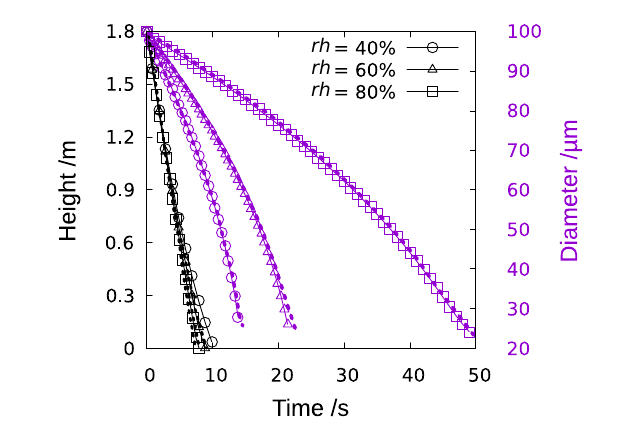}
    \caption{Validation of our model (lines) with data by Stiti et al. \cite{Stiti} (symbols).}
    \label{validation}
\end{figure}

We used the data from Stiti et al.'s \cite{Stiti} to validate the droplet evaporation and transport model. 
Figure~\ref{validation} plots the change in droplet size and falling height of a single saline droplet compared to the reference data.
The initial temperature and diameter of the droplet are 37~°C and 100~µm, respectively.
The ambient temperature is 20~°C with relative humidity of 40\%, 60\%, and 80\%.
Considering that homogeneous droplets are assumed in the reference study, we used the same assumption and the same saturation limit of $Y_{\text{2},\textup{avg}}$ = 0.357 for validation purposes. However, we used the efflorescence limit $Y_{\text {2,s}}=$~0.393 \cite{Gregson} for the rest of the paper. In this paper, $Y_{\textit{i},\textup{avg}}$ refers to the average mass fraction of component "$\textit{i}$" within the droplet and is relevant to the homogeneous droplets. On the other hand $Y_{\textit{i}}$ refers to the mass fraction of component "$\textit{i}$", for a specific spatial location within the droplet. $Y_{\textit{i},\textup{s}}$ refers to the mass fraction of component "$\textit{i}$" at the droplet surface.
Our results show good agreement with the reference data.
For the investigated conditions, a droplet with an initial diameter of 100~µm falls to the ground before reaching saturation.

\section{Results and Discussion}

In this paper, we developed an advanced saliva droplet evaporation model to predict internal concentration gradients, drying kinetics, and particle formation.
Currently, the literature does not include a precise definition of saliva composition. Also, saliva composition can be influenced by several factors, including diet, health conditions, and respiratory activity \cite{Vejerano}. On the other hand, the physicochemical properties of saliva remain limited in the available literature, posing difficulties in the direct application of existing evaporation models. For simplicity, assuming saliva as a mixture of water and NaCl is proposed by many works \cite{Xie, Wei, Li2020, Stiti}. It is a valid assumption based on saliva's composition with approximately 99\% water and 1\% various elements, primarily sodium, and chloride \cite{Almeida}. However, we noted that this assumption excludes protein, surfactant, and other components which are relevant for the physicochemical properties of the fluid and thus evaporation. For example, a surfactant accelerates evaporation by expanding the droplet radius, leading to a more rapid increase in the NaCl concentration at the droplet surface. However, the presence of a NaCl concentration gradient has the ability to trigger Marangoni flux, which pushes surfactant molecules towards areas of higher NaCl concentration. Consequently, the surfactant would generate a reversed surface tension gradient, effectively counteracting the NaCl-induced Marangoni flow and promoting uniform NaCl distribution \cite{Shao}. In order to focus on the effect of hollow particle formation on the transmission of SARS-CoV-2 we simplified saliva droplets as a bicomponent salt(NaCl)-water mixture. However, the initial NaCl mass fraction in the liquid mixture should be adjusted very carefully in order to obtain an accurate evaporation rate. Therefore, we used the initial concentration of 6 g NaCl/kg liquid mixture by Stiti et al. \cite{Stiti}, attained by evaporation experiments of acoustically levitated saliva droplets.
We computed the transport of evaporating NaCl-water droplets emerging from a cough in a flow of 0.3~m/s in different ambient conditions.
By doing so, we determine if hollow particles are formed, and whether it causes particles to travel larger distances.
We evaluate the transmission risk of SARS-CoV-2 by considering the efflorescence, virus infectivity, and virus load.
A detailed explanation of the evaporation and transport model is provided in the Methods section.

\begin{figure}[b]
\centering
\includegraphics[width=8cm]{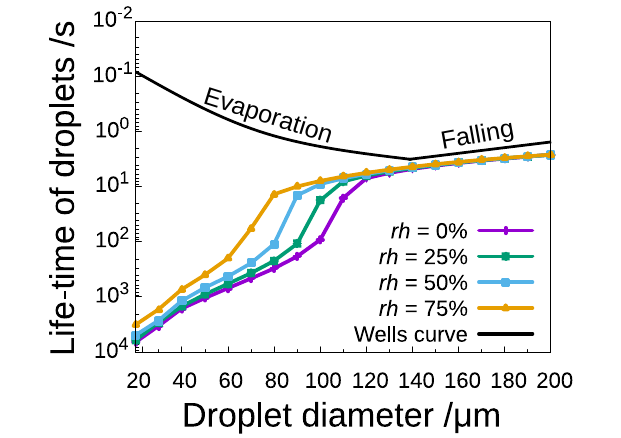}
\caption{
Contrary to the classical perception, saliva droplets do not evaporate completely but form solid nuclei.
Thus, the lifetimes of these nuclei at $T_{\text {amb}}$ = 18~°C before falling to the ground is much longer than predicted by the classical Wells curve \cite{Wells} that assumes pure water droplets.}
\label{wells}
\end{figure}

Figure~\ref{wells} compares Wells' \cite{Wells} well-known evaporation-falling curve for pure water droplets to our simulations.
All data is for a falling height of 2~m and an ambient temperature of $T_{\text {amb}}$ = 18~°C.
Wells established the evaporation part of the curve for dry and the falling part for saturated air.
The simulations also cover relative humidities of $rh=$ 0\%, 25\%, 50\%, and 75\%.

According to Wells \cite{Wells}, two different mechanisms limit the lifetime of the droplets. 
Droplets smaller than 140~µm completely evaporate, even though he speculated that nuclei might form before reaching the ground and these nuclei might float for long time periods.
Droplets larger than 140~µm fall to the ground in less than 4 seconds before evaporating.

In our simulations, the falling time of 140~µm droplets increases from 3.3 to around 4.5 seconds.
Whereas Wells assumed the droplets of the size class 140~µm to 200~µm to remain constant in size, the simulations consider their shrinkage during falling.
The settling velocity of the shrinking droplets reduces; 
thus, they take longer to reach the ground.

For droplets smaller than 140~µm, our new saliva droplet drying model provides a new view of the evaporation-falling curve:
Contrary to the evaporation falling curve of Wells, we predict that the small droplets form nuclei after drying.
The lifetime is no longer limited by complete evaporation but, like the large droplet, by their falling time to the ground.
The falling time increases with decreasing initial diameter;
for the smallest droplet considered here (20~µm), the falling time is more than one hour. 

Overall, predicting the formation of nuclei from saliva droplets changes the traditional concept of the evaporation-falling curve quantitively and qualitatively.
The falling times are longer than expected, and the droplet size class, which could be more important for the transmission of the virus, may differ from what is currently thought.

Also, Fig.~\ref{wells} indicates the weather dependency of the results: 
the effect of relative humidity on droplets larger than 140~µm is low; 
instead, their initial diameter determines their falling time.
Conversely, the falling time of small droplets becomes longer if the relative humidity decreases.
Decreasing the ambient humidity accelerates the droplet's shrinkage, allowing it to remain suspended longer.

\begin{figure}[tb]
\centering
\includegraphics[width=8cm]{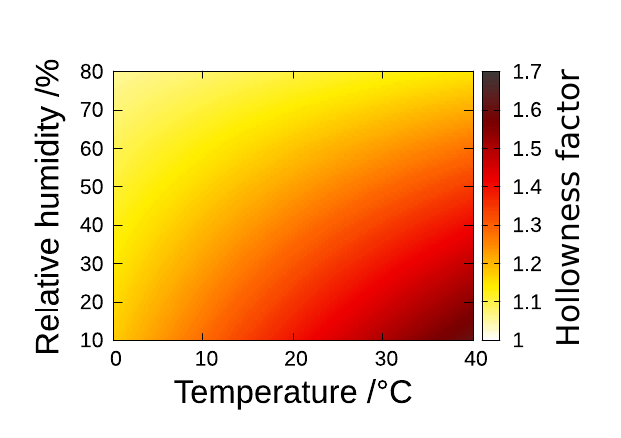}
\caption{
High ambient temperature and low humidity might lead saliva droplets to form hollow particles during evaporation because of high internal concentration gradients.
The diameter of the nuclei formed from droplets of 75~µm is 1.6 times larger than that of a droplet in cold, humid weather.} 

\label{inflation}
\end{figure}

To systematically study the weather influence, Fig.~\ref{inflation} plots the hollowness factor of saliva droplets exposed to an ambiance between $T_\mathrm{amb}=$~0~°C and 40~°C and between $rh=$~10\% and 80\%.
The hollowness factor quantifies the ratio of the volume of a hollow particle to the volume of a complete solid residue.
The hollow particle volume is fixed once the droplet surface reaches the efflorescence limit of $Y_{\text {i}}$ = 0.393 \cite{Gregson}.
The volume of the solid residue is computed assuming spatially homogeneous droplets.
The initial droplet temperature is 34~°C, which corresponds to the temperature of a saliva droplet leaving the human mouth \cite{Bourouiba2014}.
Figure~\ref{inflation} shows the hollowness factor of droplets of an initial size of 75~µm;
however, the results are qualitatively the same for the complete droplet size range covered by our study.  
The droplets were suspended until all liquid evaporated, independent of their falling height, to obtain this ratio.

The hollowness factor rises with the relative humidity decreasing and the ambient temperature increasing.
When the humidity is 80\%, the hollowness factor is at its lowest and remains below 1.100.
For $T_{\text {amb}}$~= 0~°C, 10~°C, 20~°C, 30~°C, and 40~°C, the hollowness factor rise to 1.055, 1.064, 1.072, 1.080, and 1.088, respectively.
At $rh=$~10\% and for $T_{\text {amb}}$ = 0~°C, 10~°C, 20~°C, and 30~°C, the hollowness factors are 1.178, 1.261, 1.364, and 1.481, respectively.
Within our studied conditions, the highest hollowness factor is 1.610 when the humidity is 10\% and $T_{\text {amb}} $ = 40~°C.
That means the droplet nucleus becomes 1.610 times larger in the case of formation of a hollow particle.

\begin{figure}[tb]
\centering
\includegraphics[width=12cm]{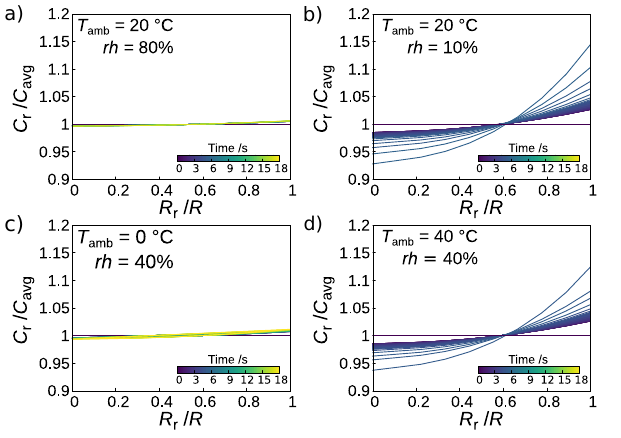}
\caption{
The solute concentration gradient within the droplet increases in a dry and warm ambiance.
The different line colors indicate the development of solute concentration with time. 
For $T_{\text {amb}} $ = 20~°C and $rh$= 80\% (a), the droplet's composition is uniform.
When the ambient humidity drops to 10\% (b), the solute concentration at the droplet's surface is 1.149 times higher than the average within the droplet.
Increasing the ambient temperature further increases the concentration gradients (see (c) and (d)).
}
\label{concgrad}
\end{figure}

Figure~\ref{concgrad} clarifies the relation of our findings to the droplet internal concentration profiles.
All simulations start with uniform species distributions;
the concentration gradients develop as the droplets evaporate.

For $rh=$ 80\% and $T_\mathrm{amb}=$~20~°C (Fig.~\ref{concgrad}a), the salt distributes almost uniformly. 
However, for $rh=$ 10\% at the same temperature (Fig.~\ref{concgrad}b), the high evaporation rate gives rise to concentration gradients.
When the droplet's surface reaches the efflorescence limit of 0.393, the concentration of the center of the droplet is 0.317, and the average concentration is 0.342.
For the same humidity but increased temperature (comparing Figs.~\ref{concgrad}c and~d), the concentration of salt at the surface of the droplet increases.
When the ambient temperature rises to $T_\mathrm{amb}=$~40~°C, the surface-to-average concentration ratio rises to 1.133.

The reason for the increased concentration gradient is that the evaporation rate outcompetes the diffusion rate.
More precisely, the solvent evaporates faster at the droplet's surface than the solute re-distributes within the droplet.
The fast evaporation enriches the surface with solutes.
A solid layer forms when the solute mass fraction at the surface eventually reaches its efflorescence limit. 
Furthermore, faster evaporation shortens the efflorescence time.
In other words, the surface concentration reaches faster its efflorescence limit.
Consequently, hollow particles are formed, and the homogeneous droplet assumption underestimates the droplet nuclei size.

\begin{figure}[tb]
\centering
\includegraphics[width=14cm]{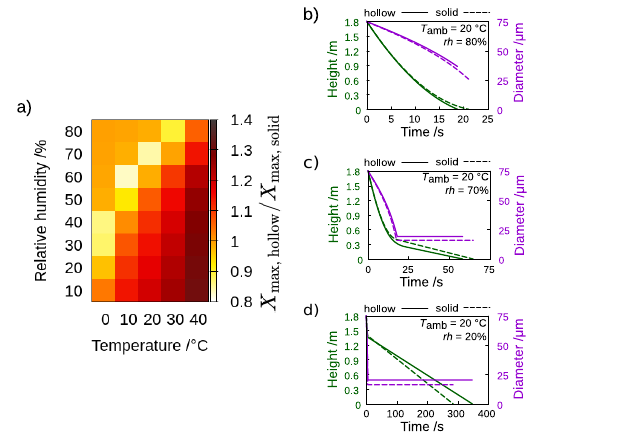}
\caption{ a) Ratio of traveling distances of a hollow and solid particle in different ambients, b) falling height and diameter of a hollow and solid particle for $T_\mathrm{amb}=$~20~°C and $rh=$ 80\%, c) $rh=$ 70\%, d) $rh=$ 20\%.
Depending on the weather conditions, hollow particles travel up to 30.4\% times further or 11\% less than solid ones (for an initial droplet size of 75~µm) (a).
When a droplet falls to the ground before a solid shell forms (e.g., for $rh=$ 80\%, $T_\mathrm{amb}=$~20~°C), the solid particles (dashed lines) travel farther than the particles (solid lines) (b).
}
\label{travelrat}

\end{figure}

The size and shape of a body determine the aerodynamic forces acting on it and, hence, its motion trajectory.
Therefore, modeling the saliva droplet size evolution and final nuclei size can provide accurate estimates of the droplet traveling distance. 
To understand the influence of the formation of a hollow particle, Fig.~\ref{travelrat}a plots the ratio of the traveling distances of a hollow and a corresponding solid particle for different ambients ($X_{\textup{max, hollow}}/X_{\textup{max, solid}}$).
In these simulations, the droplets emitted from a cough before falling to the ground from 1.8~m height, and the initial droplet size is 75~µm.
The ambient wind velocity is 0.3~m/s (The methods section gives the mathematical details of the set-up).
These conditions are exemplary but representative. 

For $rh=$ 10\% and $T_{\text {amb}}=$~40~°C the ratio is the highest (Fig.~\ref{travelrat}a); 
the hollow particles travel 1.305 times farther than the solid ones.
The ratio between both decreases with decreasing temperature; 
for 0~°C, 10~°C, 20~°C, and 30~°C the hollowness factors are 1.032, 1.139, 1.190, and 1.243, respectively.

While for most ambient conditions the hollow particles travel farther, for some combinations of humidity and temperature ($rh=$~50\% to 80\% and $T_{\text {amb}}=$~10~°C to 30~°C; $rh=$~20\% to 40\% and $T_{\text {amb}}=$~0~°C ) the relationship reverses.
When the ambient temperature is 20~°C, three scenarios occur in which the traveling distance ratio is greater, equal, or smaller than unity.
Namely, for $rh=$ 20\%, 70\%, and 80\%, the ratio becomes 1.169, 0.841, 0.982.
Thus, for some cases, the ratio drops below unity (yellow regions on Fig.~\ref{travelrat}a), which means that solid particles travel farther.

To further investigate the traveling distance ratio dropping below unity, Fig.~\ref{travelrat}b, Fig.~\ref{travelrat}c, and Fig.~\ref{travelrat}d plots the droplets' falling height and diameter with respect to time drying in an ambiance of $T_{\text {amb}}=$ 20~°C.
The solid lines represent the simulations that consider spatial gradients, in other words, droplets that solute deposit at the surface and form hollow particles, and the dashed lines represent  spatially homogeneous droplets, which form solid residues.

When $rh=$~80\%, the droplet falls to the ground before it effloresces.
Due to the droplet's short lifetime, only a little liquid mass evaporates, and the traveling distance ratio is close to unity.

Compared to $rh=$~80\%, in an ambiance of $rh=$~70\%, the droplet falls slower to the ground.
The droplet diameter evolution reaches a constant plateau at $t=19$~s, which means that a solid layer that prevents further shrinkage forms during falling.
Assuming uniform concentrations within the droplet overestimates the evaporation rate until the droplet effloresces.
Instead, the surface enriches with solutes;
thus, the vapor pressure and, consequently, the evaporation rate reduces.
Resolving the local concentration (solid curves in Fig.~\ref{travelrat}c) reveals that the droplets shrink slower and suspend shorter as anticipated before efflorescence.

When the humidity is as low as 20\%, the droplets' surface effloresces fast.
That leaves the particle enough time to completely dry out before falling to the ground.
When that happens, the volume of the hollow particle exceeds the solid particle, even though their mass is equal.
Thus, the drag force acting on the hollow particle is higher, and the particle floats longer in the air.

\begin{figure}[tb]
\centering
\includegraphics[width=8cm]{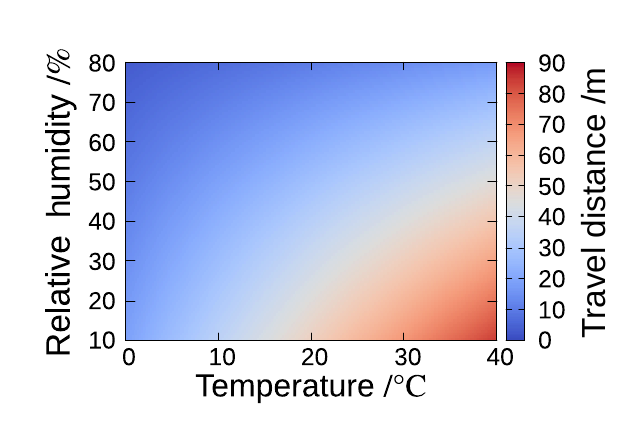}
\caption{The traveling distance of hollow particles for initial droplet diameter of 75~µm in 0.3~m/s wind increases up to 85~metres at high ambient temperature and low humidity.}
\label{travel_het}
\end{figure}

Figure~\ref{travel_het} depicts the absolute distance a droplet with 75~µm initial diameter can travel before falling to the ground from 1.8 m height in different ambients.
Again, we computed an exemplary case of a droplet emerging from a cough and suspended in a wind velocity of 0.3~m/s.
The maximum traveling distance occurs for $rh=$ 10\% and $T_{\text {amb}}=$ 40~°C, where the droplet travels 84.7~m.
The distance reduces to 74.6~m, 61.8~m, 44.0~m, and 19.0~m when $T_{\text {amb}}$ reduces to 30~°C, 20~°C, 10~°C, and 0~°C, respectively.
For a higher $rh$ of 80\% and $T_{\text {amb}}=$ 40~°C, the droplet travels only 8.3~m, and the distance reduces to 4.4~m, 3.8~m, 3.5~m, and 3.3~m for $T_{\text {amb}}=$ 30~°C, 20~°C, 10~°C, and 0~°C.

Our results demonstrate that a droplet's traveling distance increases as the relative humidity decreases and temperature increases.
Even though the values provided in Fig.~\ref{travel_het} are only valid for a wind speed of 0.3~m/s, the general conclusion that the traveling distance greatly depends on the weather conditions is independent of the wind velocity. 
\begin{figure}[tb]
\centering
\includegraphics[width=14cm]{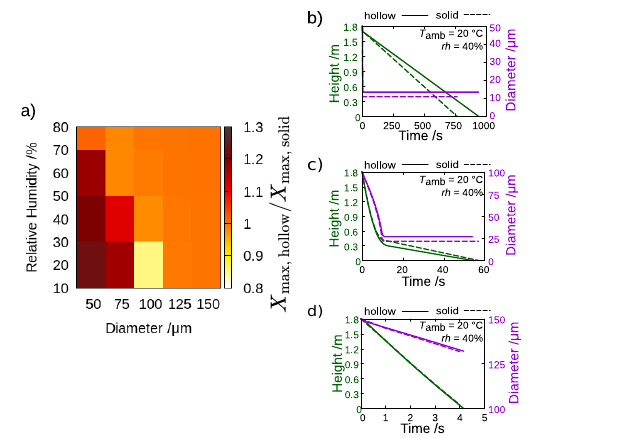}
\caption{  a) The ratio of traveling distances of a hollow and solid particle at $T_\mathrm{amb}=$~20~°C, varying ambient humidities and initial diameters b) the falling height and diameter of a hollow and solid particle at $T_\mathrm{amb}=$~20~°C, $rh=$ 40\% and for initial droplet diameter of 50 ~µm, c) 100~µm, and d) 150~µm. The formation of a hollow particle hardly affects large droplets and even decreases the traveling distance of 100~µm sized saliva droplets.
However, hollow particles travel in low humidity much further than expected.}

\label{travelrat_d}

\end{figure}

Next, we investigate the influence of the initial droplet diameter.
Figure~\ref{travelrat_d}a plots the ratio of the traveling distances of hollow and solid particles for droplets of different sizes, falling from 1.8~m at 20~°C.
For droplets with initial diameters greater than 125~µm the plotted ratio is close to unity in Figure~\ref{travelrat_d}a.
For $d_\text{p}$ = 100~µm, the ratio is below unity, namely 0.850, 0.978, 0.992, and 0.996 for $rh=$ 20\%, 40\%, 60\%, and 80\%, respectively.
When the droplets are smaller than 100~µm, their traveling distance ratio exceeds unity.
For $d_\text{p}$ = 75~µm, the ratio exceeds unity for $rh<$~50\% while it approaches unity for $rh=$~60\% to 80\%.
For $d_\text{p}$ = 50~µm, the ratio is always above one;
it increases for decreasing humidity from 1.011 ($rh=$~80\%) to 1.223 ($rh=$~20\%).

Figure~\ref{travelrat_d}b, \ref{travelrat_d}c, \ref{travelrat_d}d compares the falling height and diameter evolution of three differently sized droplets ($\textit{d}_\text{p}$ = 50~µm, 100~µm, and 150~µm) in the same ambient, $rh=$~40\% at 20~°C.
The droplet with the initial diameter of 50~µm suspends long enough to effloresce and dry. 
The predicted equilibrium size of the solid particle is 10~µm and of the hollow particle is 13~µm.
Due to the incorrect prediction of the nucleus size, current models underestimate the particle's falling time and traveling distance.

Droplets of an initial size of 100~µm fall faster to the ground than droplets of 50~µm.
The 100~µm droplets suspend long enough to effloresce but fall to the ground before they dry.
Thus, analogous to the cases in Fig.~\ref{travelrat_d}a, the homogenous simulation overpredicts the droplet's evaporation rate and traveling distance.

The droplet of the initial diameter of 150~µm falls, for the investigated ambiance, to the ground within 4 seconds.
Due to the short lifetime, the droplet barely evaporates, and both models predict nearly the same traveling distance.

Accurately predicting droplet trajectories in different ambient conditions is one cornerstone of evaluating the transmission risk of SARS-CoV-2.
However, there are also other factors that determine virus transmission, namely efflorescence \cite{Niazi, Oswin2022}, the infectivity of the virus \cite{Oswin2021}, and the virus load carried by a droplet \cite{Wolfel}.
Therefore, the critical weather condition is where droplets travel the farthest and carry the most infective virus copies. The parameters listed here are relevant to SARS-CoV-2 transmission and may differ for viruses of other types.

The effect of efflorescence in the transmission of SARS-CoV-2 is twofold:
On the one hand, the crystallization of salt protects the virus from prolonged exposure to high concentrations of salts, which may damage the viral proteins or lipid envelopes. Additionally, efflorescence facilitates the removal of residual water, resulting in a fully dried state that helps preserve the virus. \cite{Niazi}.
Conversely, Oswin et al. \cite{Oswin2022} observed rapid loss in the virus' infectivity due to the formation of efflorescence.
Despite being ambiguous, we elaborate on the droplets' falling and efflorescence time in the following.

\begin{figure}[tb]
\centering
\includegraphics[width=13cm]{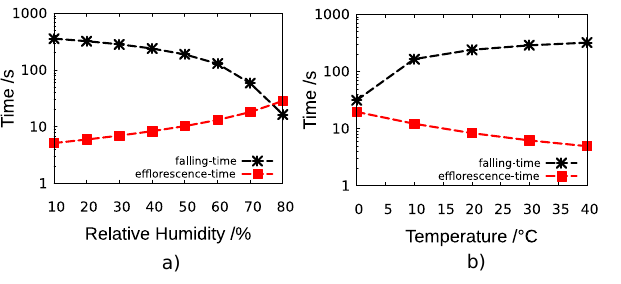}
\caption[]{
Falling and efflorescence time of droplets (75~µm) in (a) a temperature of 20~°C and varying humidity and (b) a humidity of 50\% and varying temperature.}
\label{efflorescence}
\end{figure}

Figure~\ref{efflorescence} compares the falling time (from 1.8~m) and the efflorescence time of droplets in different ambients.
The initial droplet diameter is 75~µm.
In Fig.~\ref{efflorescence}a), the ambient temperature is 20~°C, and the humidity varies, while in Fig.~\ref{efflorescence}b), the humidity is 50\%, and the temperature varies.
Our results revealed that an increase in the relative humidity and a decrease in ambient temperature increases the droplet's efflorescence time.
In contrast, the droplet's falling time decreases for the same parameter variation.

For all conditions plotted in Fig.~\ref{efflorescence}a) where the relative humidity is lower than 80\%, a droplet effloresces before falling to the ground.
Therefore, in this range, efflorescence affects airborne virus transmission.
For $rh=$ 76\%, the efflorescence and falling time are equal, meaning the droplet forms a solid shell precisely when it touches the ground. 
For $rh=$ 80\%, the droplet falls to the ground in 16 seconds while it would theoretically effloresce only after 29 seconds.
As a result, airborne transmission of SARS-CoV-2 is not affected by efflorescence in this condition.

For $rh=$ 50\%, the droplet effloresces before falling to the ground for all ambient temperatures presented in Fig.~\ref{efflorescence}b).
That means efflorescence affects airborne virus transmission over the temperatures studied here.
The difference between the falling and efflorescence time minimizes for 0~°C, in which the efflorescence time is 20 seconds and the falling time is 31 seconds and increases with increasing ambient temperature.

\begin{figure}[tb]
\centering
\includegraphics[width=15cm]{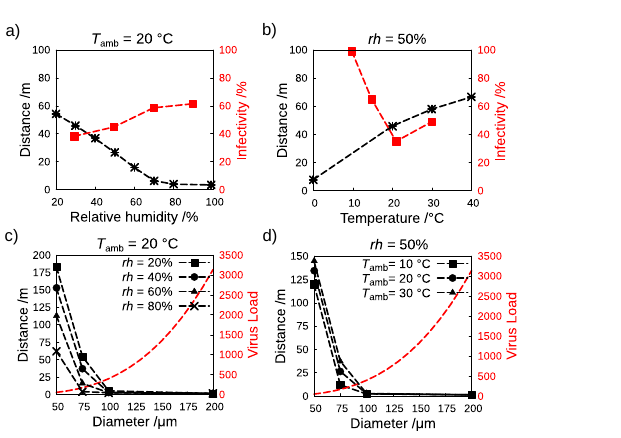}
\caption{ a) The infectivity of mouse hepatitis virus (MHV) and the distance traveled by droplets at $T_{\textup{amb}}$ = 20~°C, with different ambient humidities, and for b) $rh$ = 50\%, with different ambient temperatures. c) The load of the SARS-CoV-2 virus and the distance traveled by droplets as a function of the droplet diameter at $T_{\textup{amb}}$ = 20~°C, with different ambient humidities, and for b) $rh$ = 50\%, with different ambient temperatures. The infectivity of MHV depends on the ambient conditions and is the highest for $rh>$~60\% (a) and $T_{\text {amb}}=$~10~°C (b) \cite{Oswin2021}.
In these conditions, droplets with an initial diameter of 60~µm to 80~µm can travel far and carry many virus copies (c, d).}

\label{infect}
\end{figure}

Next, Fig.~\ref{infect} analysis the droplets' traveling distance in conjunction with the contained virus' infectivity and virus load.
Figures~\ref{infect}a) and b) plot the distance a droplet of an initial diameter of 75~µm travels in different ambients.
Additionally, it plots the infectivity of the mouse hepatitis virus (MHV) after 5 minutes of exposure to the respective ambient, as reported by Oswin et al. \cite{Oswin2021}.
The mouse hepatitis virus (MHV), an extensively studied animal coronavirus, belongs to the Betacoronavirus genus. This genus also includes SARS-CoV-2. The shared classification of MHV and SARS-CoV-2 offers an opportunity to gain valuable insights by studying MHV in relation to SARS-CoV-2 \cite{Oliveira}.
MHV's infectivity was reported to reduce to 53\% in two minutes being imposed to the environment at $rh$ = 50\% and $T_{\textup{amb}}$ = 21~°C. On the other hand, droplets of 75 ~µm suspend for much longer than 2 minutes. Therefore, a decrease in infectivity is relevant to the transmission risk of droplets with an initial size of 75~µm considered here.
It is important to highlight that in our assessment of the transmission risk, we assumed the virus's infectivity to decrease promptly. This assumption is justified by observing that the rate of infectivity decline in MHV exhibits a sharp drop within the initial seconds.

As seen in Fig.~\ref{infect}a, even though the traveling distance increases with decreasing humidity, the infectivity of MHV does not.
Contrary, the MHV' infectivity decreases to 40\% of its initial value when the humidity is 40\%.
When $rh \geq 80$, the infectivity of MHV is higher, around 60\%.
However, for this humidity, the droplet travels 3.8~m.
Thus, the virus is more infectious in high humidities, but the droplet that carries the virus can not stay airborne and travel long distances.

As regards the ambient temperature (Fig.~\ref{infect}b), the MHV is most infectious at 10~°C.
The infectivity drops rapidly below $rh=$~40\% as the ambient temperature increases to 20~°C.
The droplet at 10~°C and 20~°C travel 26.7~m and 45.8~m, respectively.
As the temperature rises to 30~°C, the infectivity rises above 40\%, and droplets travel 57.77~m.

In summary, the droplet traveling distance and the infectivity of MHV react oppositely to changing environmental conditions.
Thus, the increased transmission risk due to farther-spreading droplets might be counterbalanced by the decreased infectivity of the contained virus.
A noteworthy exception is the increase of the ambient temperature from 20~°C to 30~°C, for which both the traveling distance and the infectivity increases, thus, posing a high virus transmission risk.
The risk of transmission of SARS-CoV-2 is influenced by both the distance traveled and the number of viral copies present in a droplet, which are determined by their sizes.
Figures~\ref{infect}c) and~d) depict the traveling distances and loads (number of virus copies) of SARS-CoV-2 of droplets for different ambients and initial droplet sizes.
The plotted virus load stems from the size-dependent SARS-CoV-2 load equation by Wang et al.~\cite{Wang_virus}, the maximum virus load of an infected person of $2.35\times 10^9$ copies per ml from Wolfel et al.~\cite{Wolfel}. 
Assuming a constant viral copy density in saliva, the virus load of a droplet increases cubically with its size;
a 200~µm droplet can carry about 3000 viral copies, and a 50~µm droplet only about 48.

According to Figs.~\ref{infect}c) and~d), the droplet size has opposite effects on the traveling distance and the virus load. 
Droplets of 200~µm contain the most viruses but travel only 1.5~m, whereas droplets of 50~µm might travel up to 180~m but contain hardly any virus;
both diminishes the transmission risk of SARS-CoV-2.

The data presented in this paper allows us to evaluate the transmission risk of SARS-CoV-2 in various ambient conditions.
Depending on the ambient humidity, a droplet of an initial size of 75~µm travels up to 60~m and contains a relatively high virus load (Fig.~\ref{infect}c).
More specifically, droplets with diameters of 60~µm, 70~µm, and 80~µm carry up to 84, 134, and 200 viral copies, respectively.
Further, in an ambient humidity greater than 60\%, where the virus is highly infectious, droplets of a size of 60~µm to 80~µm still travel large distances.
Thus, droplets of this size can carry at $T_\mathrm{amb}=20$~°C many highly infectious viral copies over long distances (Fig.~\ref{infect}c).
When $rh$ = 50\%, the infectivity of the virus is the highest for $T_\mathrm{amb}=10$~°C while the infectivity remains 50\% for $T_\mathrm{amb}=30$~°C  (Fig.~\ref{infect}d).
For both $T_\mathrm{amb}=10$~°C and $T_\mathrm{amb}=30$~°C, droplets in the range 60~µm to 80~µm can travel long distances and carry a higher number of viruses.

\section{Conclusions}
The key finding of our study is that, after exhaling, dried saliva (NaCl-water) droplets can form hollow particles (Fig.~\ref{inflation}), which alters the particles' trajectories (Fig.~\ref{travelrat}).
Overall, the size evolution and trajectories of saliva droplets are imprecisely predicted by current models that neglect the non-uniform distribution of non-volatiles within a droplet (Fig.~\ref{wells}).
For an ambient humidity below 80\%, internal concentration gradients lead to hollow particle formation.
Particles' hollowness increases more in hot and dry environments and less in cold and moist environments.

In dry and cold ambients (Fig.~\ref{travelrat}) or if the droplet is large (Fig.~\ref{travelrat_d}), droplets fall to the ground before or shortly after a shell forms on their surface.
Then, the droplets have a shorter airborne lifetime than known so far.
However, droplets that suspend long enough to effloresce and dry, suspend longer and travel farther.
Due to hollow particle formation, a 75~µm sized droplet exposed to an ambient humidity of 10\% and temperature of 40~°C and a wind velocity of 0.3~m/s can travel up to 84.7~m (Fig.~\ref{travel_het}).
However, it falls to the ground within 5~m when the humidity is 80\%.

The transmission risk of SARS-CoV-2 not only depends on the saliva droplets' falling time and traveling distance, but are also sensitive to changes in temperature, humidity, and other environmental factors, and these conditions can affect their ability to survive and spread. 
In dry and cold environments, droplets suspend long, but the virus is less infectious. 
Contrary, in humid and warm environments, the virus is more infectious, but the droplets fall fast to the ground (Fig.~\ref{infect}a, b).
Further, large droplets travel less far but carry more SARS-CoV-2 copies, while small droplets can travel farther but contain fewer viruses (Fig.~\ref{infect}c, d).
Given that saliva droplets might form hollow particles and travel farther than previously known, it suggests that facial masks may offer a higher level of protection against SARS-CoV-2 compared to solely relying on social distancing measures.
Further, this study supports that the transmission of SARS-CoV-2 is complex, depending on many physical and biological mechanisms and their interactions. 
In line with our inferences, static social distancing rules meant to be valid for every situation are misleading since the ambiance strongly affects the transmission risk.
Assessing potential transmission risks and tailoring regulations requires sophisticated mathematical models, to which we contribute with this work.
Then, these models can be part of assessing the virus transmission risk in specific situations, in and outdoors, and the effectiveness of protective measures.

\section*{Acknowledgements}
Funded by the Deutsche Forschungsgemeinschaft (DFG, German Research Foundation) – Projektnummer 468822780.


\begin{thebibliography}{99}
\bibitem {WHO} World Health Organization. $\textit{Modes of transmission of virus causing COVID-19: implications}$
$\textit{for IPC precaution recommendations}$. (2020).
\bibitem {Brankston} Brankston, G., Gitterman, L., Hirji, Z., Lemieux, C. and Gardam, M.\, Transmission of influenza A in human beings. $\textit{Lancet Infect. Dis.}$ $\mathbf7$, 257-265 (2007).
\bibitem {Goldman} Goldman, E.\, Exaggerated risk of transmission of COVID-19 by fomites. $\textit{Lancet Infect. Dis.}$ $\mathbf{20}$, 892-893 (2020).
\bibitem {Lewis} Lewis, D.\, COVID-19 rarely spreads through surfaces. So why are we still deep cleaning? $\textit{Nature}$ $\mathbf{590}$, 26-28 (2021).
\bibitem {Morawska} Morawska, L.\, Droplet fate in indoor environments, or can we prevent the spread of infection? $\textit{Indoor Air}$ $\mathbf{16}$, 335-47 (2006).
\bibitem {Zhang} Zhang, R., Li, Y., Zhang, A., Wang, Y. and Molina, M.\, Identifying airborne transmission as the dominant route for the spread of covid-19. $\textit{Proc. Natl. Acad. Sci. U.S.A.}$ $\mathbf{117}$, 14857-14863 (2020).
\bibitem {Wells} Wells, W. F. On air-borne infection: study II. Droplets and droplet nuclei. $\textit{Am. J. Epidemiol.}$ $\mathbf{20}$, 611-618 (1934).
\bibitem {WHO2} World Health Organization. $\textit{Management of ill travellers at points of entry – international }$ 
$\textit{airports, ports and ground crossings – in the context of COVID-19 outbreak}$. (2020).
\bibitem {Thakur} Thakur, A. K., Kaundle, B. and Singh, I. Mucoadhesive Drug Delivery Systems in Respiratory Diseases. in $\textit{Targeting Chronic Inflammatory Lung Diseases Using Advanced Drug Delivery}$
$\textit{Systems}$ 475-491 (Academic Press, 2020).
\bibitem {Ahmadzadeh} Ahmadzadeh, M., Farokhi, E. and Shams, M. Investigating the effect of air conditioning on the distribution and transmission of COVID-19 virus particles. $\textit{J. Clean Prod.}$ $\mathbf{316}$, 128-147 (2021).
\bibitem {Almeida} Almeida, P., Grégio, A., Machado, M., Lima, A. and Azevedo, L. Saliva composition and functions: A comprehensive review. $\textit{J. Clean Prod.}$ $\mathbf{9}$, 72–80 (2008).
\bibitem {Czerwiec} Czerwiec, T. et al. Thermal management of metallic surfaces: evaporation of sessile water droplets on polished and patterned stainless steel. $\textit{IOP Conf. Ser.: Mater. Sci. Eng.}$ $\mathbf{258}$, 012003 (2017).
\bibitem {Liu} Liu, Z., et al. Single fiber optical trapping of a liquid droplet and its application in microresonator.  $\textit{Opt. Commun.}$ $\mathbf{381}$, 371–376 (2016).
\bibitem {Chaudhuri} Chaudhuri, S., Basu, S., Kabi, P., Rajasekharan Unni, V. and Saha, A. Modeling the role of respiratory droplets in covid-19 type pandemics. $\textit{Phys. Fluids}$ $\mathbf{32}$, 063309 (2020).
\bibitem {Basu} Basu, S., Kabi, P., Chaudhuri, S. and Saha, A. Insights on drying and precipitation dynamics of respiratory droplets from the perspective of COVID-19. $\textit{Phys. Fluids}$ $\mathbf{32}$, 123317 (2020).
\bibitem {Stiti} Stiti, M., Castanet, G., Corber, A., Aldén, M. and Berrocal, E. Transition from saliva droplets to solid aerosols in the context of COVID-19 spreading. $\textit{Environ. Res.}$ $\mathbf{204}$, 112072 (2021)
\bibitem {Lieber} Lieber, C., Melekidis, S., Koch, R. and Bauer, H. J. Insights into the evaporation characteristics of saliva droplets and aerosols: Levitation experiments and numerical modeling. $\textit{J. Aerosol Sci.}$ $\mathbf{154}$, 105760 (2021).
\bibitem {Jayaweera} Jayaweera, M., Perera, H., Gunawardana, B. and Manatunge, J. Transmission of COVID-19 virus by droplets and aerosols: A critical review on the unresolved dichotomy. $\textit{Environ. Res.}$ $\mathbf{188}$, 109819 (2020).
\bibitem {Bourouiba} Bourouiba, L. Turbulent gas clouds and respiratory pathogen emissions: potential implications for reducing transmission of COVID-19. $\textit{JAMA}$ $\mathbf{323}$, 1837–1838 (2020).
\bibitem {Xie} Xie, X., Li, Y., Chwang, A. T., Ho, P. L. and Seto, W. H. How far droplets can move in indoor environments - revisiting the Wells evaporation-falling curve. $\textit{Indoor Air}$ $\mathbf{17}$, 211-225 (2007).
\bibitem {Speight} Speight, J. G. Sources and Types of Organic Pollutants. in $\textit{Environmental Organic}$ $\textit{Chemistry for Engineers}$ 153-201 (Elsevier, 2017).
\bibitem {Ugarte-Anero} Ugarte-Anero, A., Fernandez-Gamiz, U., Portal-Porras, K., Zulueta, E. and Urbina, O. Computational characterization of the behavior of a saliva droplet in a social environment. $\textit{Sci. Rep.}$ $\mathbf{12}$, 6405 (2022).
\bibitem {Chillon} Chillon, S., Ugarte-Anero, A., Iradi, I., Fernandez-Gamiz, U. and Zulueta, E., 2021. Numerical modeling of the spread of cough saliva droplets in a calm confined space. $\textit{Mathematics}$ $\mathbf{9}$, 574 (2021).
\bibitem {Biswas} Biswas, R., Pal, A., Pal, R., Sarkar, S. and Mukhopadhyay, A. Risk assessment of covid infection by respiratory droplets from cough for various ventilation scenarios inside an elevator: An OpenFOAM-based computational fluid dynamics analysis. $\textit{Phys. Fluids}$ $\mathbf{34}$, 013318 (2022).
\bibitem {Gregson} Gregson, F. K. A., Robinson, J. F., Miles, R. E.H., Royal, C. P. and Reid, J. P. Drying kinetics of salt solution droplets: Water evaporation rates and crystallization. $\textit{J. Phys. Chem.  B}$ $\mathbf{123}$, 266-276 (2018). 
\bibitem {Rezaei} Rezaei, M. and  Netz, R. R. Water evaporation from solute-containing aerosol droplets: Effects of internal concentration and diffusivity profiles and onset of crust formation. $\textit{Phys. Fluids}$ $\mathbf{33}$, 091901 (2021). 
\bibitem {Nesic} Nešić, S., and Vodnik, J.  Kinetics of droplet evaporation. $\textit{Chem. Eng. Sci.}$ $\mathbf{46}$, 527-537 (1991).
\bibitem {Boel} Boel, E., Koekoekx, R., Dedroog, S., Babkin, I., Vetrano, M. R., Clasen, C., and Van den Mooter, G. Unraveling particle formation: From single droplet drying to spray drying and electrospraying. $\textit{Pharmaceutics}$, $\mathbf{12}$, 625 (2020).
\bibitem {Walton} Walton, D. E. and Mumford, C. J. The morphology of spray dried particles The effect of process variables upon the morphology of spray-dried particles. $\textit{Trans. Inst. Chem. Eng.}$ $\mathbf{77A}$ 442–460 (1999).
\bibitem {Hardy} Hardy, D. A., Robinson, J. F., Hilditch, T. G., Neal, E., Lemaitre, P., Walker, J. S., and Reid, J. P. Accurate Measurements and Simulations of the Evaporation and Trajectories of Individual Solution Droplets. $\textit{J. Phys. Chem. B .}$ $\mathbf{15}$, 3416-3430 (2023).
\bibitem {Zhang} Zhang, J., Zhang, S., Wang, Z., Zhang, Z., Wang, S., and Wang, S. Hopper‐Like Single Crystals of Sodium Chloride Grown at the Interface of Metastable Water Droplets. $\textit{Angew. Chem., Int. Ed.}$ $\mathbf{50}$, 6044-6047 (2011). 
\bibitem {Vehring} Vehring, R. Pharmaceutical particle engineering via spray drying. $\textit{CPharm. Res.}$, $\mathbf{25}$, 999-1022 (2008).
\bibitem {Vicente} Vicente, J., Pinto, J., Menezes, J. and Gaspar, F. Fundamental analysis of particle formation in spray drying.  $\textit{Powder Technol.}$  $\mathbf{247}$, 1–7 (2013).
\bibitem {Lucas} Chegini, G. R. and Ghobadian, B. Effect of Spray-drying Conditions on Physical Properties of Orange Juice. $\textit{ Powder. Dry. Tech.}$  $\mathbf{23}$, 657–668 (2005).
\bibitem {Oswin2021} Oswin, H. P. et al. Measuring stability of virus in aerosols under varying environmental conditions. $\textit{Aerosol Sci. Technol.}$ $\mathbf{55}$, 1315-1320 (2021).
\bibitem {Wang_virus} Wang, Y., Xu, G. and Huang, Y. W.  Modeling the load of SARS-CoV-2 virus in human expelled particles during coughing and speaking. $\textit{PLoS ONE}$ $\mathbf{15}$, 0241539 (2020).
\bibitem {Gopireddy} Gopireddy, S. R. and Gutheil, E. Numerical simulation of evaporation and drying of a bi-component droplet. $\textit{Int. J. Heat Mass Transf.}$ $\mathbf{66}$, 404-411 (2013).
\bibitem {Law} Law, C. K. Recent advances in droplet vaporization and combustion. $\textit{PECS}$ $\mathbf{8}$, 171-201 (1982).
\bibitem {Abramzon} Abramzon, B.,  and  Sirignano, W. A.  Droplet vaporization model for spray combustion calculations. $\textit{ICHMT}$ $\mathbf{32}$, 1605-1618 (1989).
\bibitem {Hubbard} Hubbard, G. L., Denny, V. E., and Mills, A. F. (1975). Droplet evaporation: effects of transients and variable properties. International journal of heat and mass transfer, 18(9), 1003-1008.
\bibitem {Gopireddy2} Gopireddy, S. R., and E. Gutheil. "Modeling and Simulation of Water Evaporation from a Droplet of Polyvinylpyrrolidone (PVP) Aqueous Solution."  $\textit{Proceedings of the ICLASS 2012, 12th Triennial International Conference on Liquid atomi-}$ $\textit{zation and Spray Systems, Heidelberg, Germany, 2-6 September 2012.}$
\bibitem {Gong} Gong, N. et al. The acceptable air velocity range for local air movement in the tropics, $\textit{HVACandR Res.}$ $\mathbf{15}$, 1065–1076 (2006).
\bibitem {Putnam} Putnam, A. Integratable form of droplet drag coefficient, $\textit{J. Am. Rocket Soc.}$, 1467-4798 (1961).
\bibitem {Wang_cough} Wang, H., Li, Z., Zhang, X., Zhu, L., Liu, Y. and Wang, S. The motion of respiratory droplets produced by coughing. $\textit{Phys. Fluids}$ $\mathbf{32}$, 125102 (2020).
\bibitem {Vejerano} Vejerano, E. P., and Marr, L. C. Physico-chemical characteristics of evaporating respiratory fluid droplets. $\textit{J. R. Soc. Interface}$ $\mathbf{15}$, 20170939 (2018).
\bibitem {Wei} Wei, J., and Li, Y. Enhanced spread of expiratory droplets by turbulence in a cough jet. $\textit{Build Environ.}$ $\mathbf{93}$, 86-96 (2015).
\bibitem {Li2020} Li, H., Leong, F. Y., Xu, G., Ge, Z., Kang, C. W., and Lim, K. H. Dispersion of evaporating cough droplets in tropical outdoor environment. $\textit{Phys. Fluids}$ $\mathbf{32}$, 113301 (2020).
\bibitem {Shao} Shao, X., Hou, Y., and Zhong, X. Modulation of evaporation-affected crystal motion in a drying droplet by saline and surfactant concentrations. $\textit{Colloids Surf. A: Physicochem. Eng.}$, $\mathbf{623}$, 126701 (2021).
\bibitem {Bourouiba2014} Bourouiba, L., Dehandschoewercker, E. and Bush, J. W. M. Violent expiratory events: On coughing and sneezing. $\textit{J. Fluid Mech.}$ $\mathbf{745}$, 537–563 (2014). 
\bibitem {Niazi} Niazi, S., Short, K. R., Groth, R., Cravigan, L., Spann, K., Ristovski, Z. and Johnson, G. R. Humidity-dependent survival of an airborne influenza A virus: Practical implications for controlling airborne viruses. $\textit{Environ. Sci. Technol. Lett.}$ $\mathbf{8}$, 412–418 (2021).
\bibitem {Oswin2022} Oswin, H. P. et al. The dynamics of SARS-CoV-2 infectivity with changes in aerosol microenvironment. $\textit{Proc. Natl. Acad. Sci.(USA)}$ $\mathbf{119}$, e2200109119 (2022).
\bibitem {Wolfel} W\"olfel, R. et al. Virological assessment of hospitalized patients with COVID-2019. $\textit{Nature}$ $\mathbf{581}$, 465–469 (2020).
\bibitem {Oliveira} Oliveira, G. P., and Kroon, E. G. Mouse hepatitis virus: A betacoronavirus model to study the virucidal activity of air disinfection equipment on surface contamination. $\textit{J. Virol. Methods}$ $\mathbf{297}$, 114274 (2021).
\end{thebibliography}
\end{document}